\documentclass[useAMS,usenatbib]{mn2e}

\usepackage{graphicx} 
\newcommand{\Stromgren}{Str\"omgren\ }

\begin{document}
\title[On distinguishing age from metallicity with photometric data ]
{On distinguishing age from metallicity with photometric data }
\author[Baitian Tang and Guy Worthey ]{Baitian Tang$^{1}$\thanks{E-mail:
baitian.tang@email.wsu.edu (BT)} 
and Guy Worthey$^{1}$\thanks{E-mail:
gworthey@wsu.edu (GW)}\\
$^{1}$Department of Physics and Astronomy, Washington State
University, Pullman, WA 99163-2814, USA\\}

\date{Accepted  . Received  ; in original form 2011}

\pagerange{\pageref{firstpage}--\pageref{lastpage}} \pubyear{ }

\maketitle

\label{firstpage}

\begin{abstract}
In the study of galaxy integrated light, if photometric indicators
could extract age and metallicity information of high enough quality,
photometry might be vastly more efficient than spectroscopy for the
same astrophysical goals. Toward this end, we search three photometric systems: David Dunlap
Observatory (DDO), Beijing-Arizona-Taiwan-Connecticut
(BATC), and \Stromgren systems for their ability to disentangle age and
abundance effects. Only the
\Stromgren $[c_{1}]$ vs. $[m_{1}]$ plot shows moderate age-metallicity
disentanglement. We also add to the discussion of 
optical to near-infrared Johnson-Cousins broad band colours, 
finding a great decrease in age sensitivity when updated isochrones
are used.
\end{abstract}


\section{Introduction}
\label{sect:intro}
Since the concept of stellar populations was invented \citep{ba44},
stellar population synthesis (SPS) has been proven to be a useful tool
for revealing clues about galaxy formation \citep{tin68, tin78,
  bc93}. Research on several crucial factors of stellar modelling,
like convection, opacity, heavy-element mixing, helium content, and
mass loss \citep{char96}, resulted in increasingly accurate SPS models
\citep{bc93, w94, b94, bc03, d08, b08, b09}. In this scheme,
estimation of single burst equivalent age and metallicity, was
accomplished by comparing integrated light of observed galaxy with
that of SPS models.  However, most of the colours and absorption
feature strength appear identical if the changes of age and
metallicity satisfy $\delta \log{(\rm age\/)} \approx -3/2\; \delta
\log{(Z)}$. This so-called age-metallicity degeneracy \citep{w94}
blocks the way to estimating galaxy accurate ages from optical broad
band colours. At spectroscopically narrow bands, however, several age
sensitive or metallicity sensitive Lick indices \citep{bur84, fab85,
  g93, w942, wo97, TMB03}, taken in pairs, are effective at breaking
this degeneracy.  Balmer indices vs. Fe-peak indices are widely used,
for example, because they are relatively easy to observe (all features
are in the optical so one spectrograph can cover them all) and the
model grids in the observed index-index space open up to nearly
orthogonal grids rather than collapsed into linear, degenerate
overlapping segments.

Spectroscopy is effective, but photometry also has its strong
points. It is much more efficient for faint objects and low surface
brightness galaxies.  In the epoch of large sky surveys, like {\it
  Sloan Digital Sky Survey (SDSS), Two Micron All Sky Survey (2MASS),
  Spitzer Space Telescope (SST), Large Synoptic Survey Telescope
  (LSST)\/}, and {\it James Webb Space Telescope (JWST)\/}, photometry
is undoubtedly more feasible than spectrometry.  In fact, optical
near-infrared (near-IR) colours have been exploited to possibly
constrain age and metallicity by several studies \citep{pe90, de96,
  bell00, car09, con09}. \citet{car03} suggested ($V-K$) colour shows
good balance between parameter degeneracy and
sensitivity. \citet{pe08} found that four age intervals ($>$10 Gyr,
2$-$9 Gyr, 1$-$2 Gyr, and 0.2$-$1 Gyr) are clearly separated in the
optical near-IR colour-colour plot. These results hinge upon the
models, and specifically upon the number of thermally pulsing
asymptotic giant branch (TP-AGB) stars, which dominate near-IR light
between 0.3 and 2 Gyr of age \citep{m05, lee07}.

The present work looks for age-sensitive or metallicity-sensitive
indices in three photometric systems: David Dunlap Observatory (DDO,
\citealt{mv68,mc76}), Beijing-Arizona-Taiwan-Connecticut (BATC,
\citealt{fan96, shang98}), and \Stromgren \citep{str66,ra91, ra96}
systems. These filters are narrower in wavelength than Johnson-Cousins
broad band filters, but resemble the {\it NIRCam\/} narrow
filters\footnote{http://ircamera.as.arizona.edu/nircam/} on board {\it
  JWST\/}.  Can we find indices which are practical to break the
age-metallicity degeneracy?  The tables and graphs in $\S$
\ref{sub:three} explore this issue. In addition, equipped with the
latest stellar evolutionary isochrones of the Padova
group\footnote{http://stev.oapd.inaf.it/YZVAR/} (\citealt{b08, b09},
hereafter B09), we plot the model grids in ($B-V$) vs. ($V-K$) space,
and compare with previous models in $\S$ \ref{sub:bvk}. Observables
from three samples are chosen to verify the credibility of parameters
given by the model grids. Discrepancies between two model grids are
briefly discussed in $\S$ \ref{sect:disc}, and a brief summary of the
results and outlook for the future is given in $\S$ \ref{sect:sum}.

\section{Analysis}
\label{sect:sensi}

\subsection{DDO, BATC \& \Stromgren}
\label{sub:three}

In order to find the age-sensitive or metallicity-sensitive index, we
calculate the the metallicity-to-age sensitivity parameter ($Z _{sp}$)
using the evolving \citet[hereafter W94]{w94} models, following
\citet{w94} and \citet{ser11}. The spectral library is synthetic and is sensitive to individual elemental abundances in the 300 to 1000 nm wavelength range. The metallicity sensitivity parameter is:
\begin{equation}
Z_{sp}=\frac{\delta I_{m}/\delta \log{(Z)}}{\delta I_{a}/\delta \log{(\rm age\/)}}
\end{equation}
Where $\delta I_{m}/\delta \log{(Z)}$ is the metallicity partial
derivative of the index at 8 Gyr, and $\delta I_{a}/\delta
\log{(\rm age\/)}$ is the age partial derivative of the index at
solar metallicity.  Thus, $Z_{sp}$ is the metallicity sensitivity
at 8 Gyr, solar metallicity.

$Z_{sp}$ is surveyed for indices in three photometric
systems\footnote{ http://astro.wsu.edu/models/isochrones.html}. We
refer the readers to related papers for detailed description of
filters and indices (See $\S$\ref{sect:intro}).
As shown in \citet{w94}, indices with $Z_{sp}$ close to 1.5 are less
likely to break the age-metallicity degeneracy. Hence, indices with
extreme $Z_{sp}$ values are selected as candidates and listed in Table
\ref{tab:ss}\footnote{ Since the K band is suggested to be sensitive to
  young stellar populations,
  nonstandard colours relative to the K band, like DDO41$-$K, BATC6075$-$K, were
  also examined. However, no promising index is found. }.
What is not quantified in Table \ref{tab:ss} is the dynamic range
(compared to observational error) of a given index, 
and the dynamic range is low in DDO C(41-42), alas.

\begin{table}
\centering
\caption{Metallicity sensitivities at 8 Gyr, solar metallicity, 
for our best candidate pairs of colors} \label{tab:ss}
\begin{tabular}{rr}
\hline
Index & $Z_{sp}$\\
\hline
DDO C(41-42) & -0.60 \\
DDO C(35-48) & 1.42 \\
BATC 6075-6660 &1.15 \\
BATC 5795-6075& 1.97 \\
\Stromgren $[m_{1}]$ &1.23 \\
\Stromgren $[c_{1}]$  &2.41 \\
\hline
\end{tabular}
\end{table}

\begin{itemize}
\item{DDO system}\\ The extremely low $Z_{sp}$ value of C(41-42)
  colour makes it a promising age indicator to break the degeneracy.
  Figure \ref{fig:ddo1} shows C(35-38) vs. C(41-42) plot of our
  models using W94 stellar evolution. The isochrones are well
  separated, except SSPs with very low metallicities. However, the
  major problem is small dynamic range ($\approx$0.05 mag) of
  C(41-42). A little observational error or dust content (A$_V=1.0$
  mag is sketched in Figure \ref{fig:ddo1} to estimate the
  extinction effect) will destroy
  the well spaced isochrones.  The short distance between the central
  wavelengths of DDO 41 and 42 filters ($\approx$100 \AA) is the
  reason for this defect.


\begin{figure}
\centering
\includegraphics [scale=0.4]{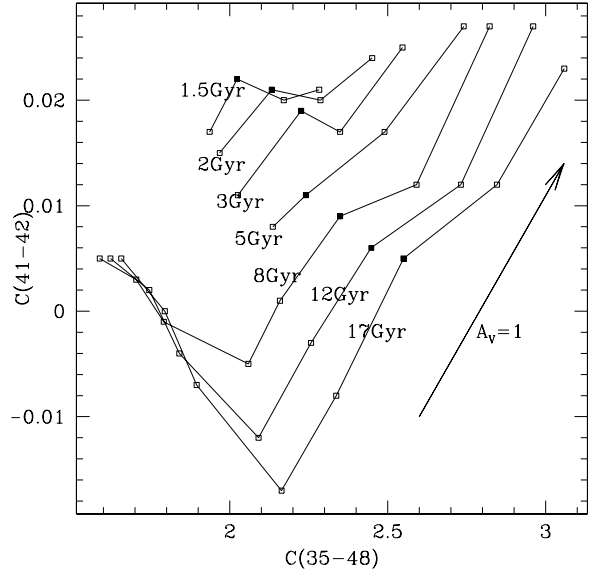}
\caption{DDO C(35-48) vs. C(41-42) plot. SSPs of the same age
  are connected as isochrones. Our models (W94) are given at age$=1.5, 2, 3,
  5$ with $\log{(Z)}=-0.225, 0, 0.25, 0.50$, and age$=8, 12, 17$ Gyr
  with $\log{(Z)}=-2.0, -1.5, -1.0, -0.50, -0.25, 0, 0.25, 0.50$. Solar
  metallicity SSPs are marked as {\it filled squares\/} to guide the
  eye. A vector for A$_V=1.0$ mag is sketched to estimate the
  extinction effect.
}\label{fig:ddo1}
\end{figure}

\begin{figure}
\centering
\includegraphics [scale=0.4]{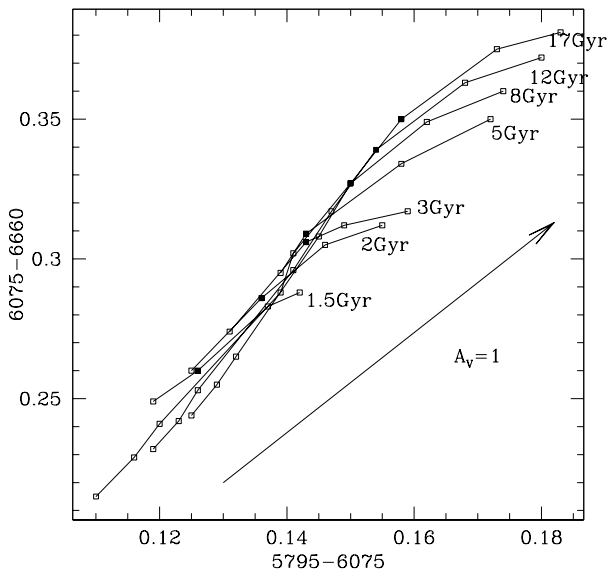} 
\caption{BATC 5795-6075 vs. 6075-6660 plot. Isochrones are
  described in Figure \ref{fig:ddo1}. A vector for A$_V=1.0$ mag is
  also sketched to estimate the extinction effect.}\label{fig:batc}
\end{figure}


\item{BATC system}\\
BATC 6075-6660 and 5795-6075 colours show potential of breaking the
degeneracy for their $Z_{sp}$ values. However, Figure
\ref{fig:batc} does not work out as expected. Though the super-solar
metallicity SSP isochrones are separated as indicated by
$Z_{sp}$, the sub-solar metallicity SSP isochrones are still highly
degenerate. We note that the 6660 filter contains H$\alpha$, which
explains age sensitivity of the 6075-6660 colour. However, we also note
that the dynamic range of both colours is small compared to the
A$_V=1.0$ mag extinction vector, and that the 6075-6660 colour only
spans 0.06 mag total, which seems small compared to reasonable
observational errors. Furthermore, since it is H$\alpha$ that is
driving the age sensitivity of this colour, it will be sensitive also
to nebular emission, which can easily overwhelm the stellar absorption
in many galaxies.

\begin{figure}
\centering
\includegraphics [scale=0.4]{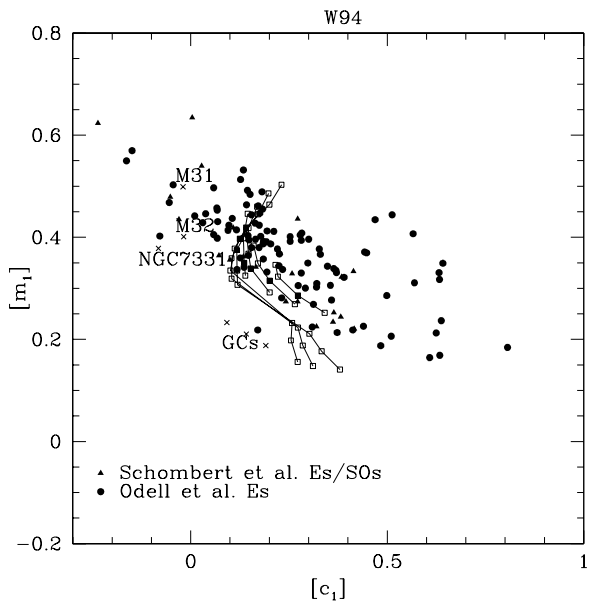} 
\caption{\Stromgren $c_{1}$ vs. $m_{1}$ plot. Isochrones are described
  in Figure \ref{fig:ddo1}. Ages are not labelled
  for narrow space, but they can be  easily inferred from Figure
  \ref{fig:mc11}. Observables from three samples are labelled
with different symbols: {\it Filled Triangles\/}: Es and S0s from
\citet{sch93}. {\it Crosses\/}: Nearby Es and GCs from
\citet{ra90}. {\it Filled Circles\/}: Es from \citet{o02}. 
}\label{fig:mceg}
\end{figure}

\begin{figure}
\centering
\includegraphics [scale=0.4]{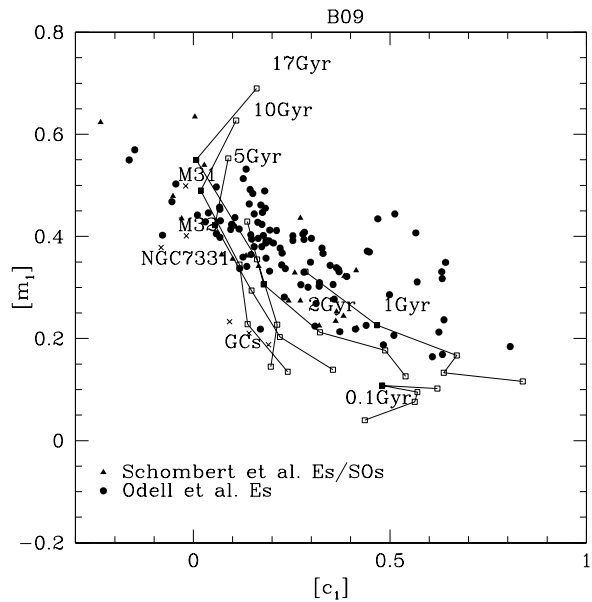} 
\caption{\Stromgren $c_{1}$ vs. $m_{1}$ plot of B09 models. The models are given
  at age$=0.1, 1, 2, 5, 10, 17$ Gyr with $\log{(Z)}=-2.23, -1.23, -0.53,
0, 0.37$. Solar metallicity SSPs are marked as {\it filled squares\/} to guide
the eye. Observable symbols are the same as in Figure \ref{fig:mceg}. }\label{fig:mc11}
\end{figure}

\item{\Stromgren system}\\
Since $[c_{1}]$ and $[m_{1}]$\footnote{$c_{1} \equiv
(u-v)-(v-b)$, $m_{1} \equiv (v-b)-(b-y) $; \\
$[c_{1}] \equiv c_{1}-0.2(b-y)$,
$[m_{1}] \equiv m_{1}+0.18(b-y)$. } are reddening
independent \citep{str66}, and they also have promising $Z_{sp}$ values, we
take a close look at these two indices. 
\citet{ra91} suggested $m_{1}$ is a metal-line idex, while $c_{1}$ is a Balmer
discontinuity index. 
Figure \ref{fig:mceg} confirms $[m_{1}]$ as a metallicity sensitive
index, because the SSP isochrones lie almost parallel to the $[m_{1}]$ axis.
With its reasonable dynamic range ($\approx$0.6 mag) and its reddening
free advantage, 
$[c_{1}]$ vs. $[m_{1}]$ plot is a promising candidate index that might break the age-metallicity
degeneracy. Three samples are  selected to test the competency of the
$[c_{1}]$ vs. $[m_{1}]$ plot: (1) Ellipticals (Es) and S0s from \citet{sch93}, shown as
{\it filled Triangles\/} in Figure \ref{fig:mceg}; (2) Nearby Es and globular clusters
(GCs) from \citet{ra90}. They are labelled separately for
clarity ({\it crosses\/});
(3) Es from \citet{o02}, shown as {\it filled circles\/}. Note that the last two
samples are recorded in a modified \Stromgren filter system called {\it uz, vz, bz, yz\/}. 
The relations of \citet{ra96} are employed to convert the observables
to the {\it uvby\/} system.  
Figure \ref{fig:mceg} shows that our model grids based on W94
evolution do not fit the observations well. B09 model grids are also
plotted along the observables in Figure \ref{fig:mc11}. Comparing with
Figure \ref{fig:mceg}, the coverage
improves greatly in the later model grids, but isochrones of age
between 5 and 17 Gyr are still partially degenerate. Obviously, the $[m_{1}]$
indices of old metal-rich SSPs in B09 models are systematically larger
than that in W94 models. 
We also note that two sets of tracks are not the same, e.g., young age, low
metallicity SSPs are not included in W94 models. To investigate the
topic furthermore , we draw the $[c_{1}]$ vs. $[m_{1}]$ plot of B94 models
($\S$ \ref{sub:bvk}), which looks strikingly similar to Figure
\ref{fig:mc11} (not shown in this paper). Thus, the stellar phases causing the dramatic
discrepancies between Figure \ref{fig:mceg} and \ref{fig:mc11} must be
different in W94 and B09 models, but similar in B94 and B09 models.
We defer discussion of the specific stellar phases because it
is beyond the scope of this work.

\end{itemize}

\subsection{Element Sensitivity}
\label{sub:sen}

\begin{table}\scriptsize
\centering
\caption{Index changes at 8 Gyr, solar metallicity
} \label{tab:eh}
\begin{tabular}{rrrrrrr}
\hline
Index & C & N & O & Na & Mg & Fe\\
\hline
DDO C(41-42) &     0.028  & 0.039 & -0.018 &  -0.001 &  -0.002 & -0.001 \\
DDO C(35-48) &      0.068 &  0.026 &  -0.033 & -0.005 & -0.002  & 0.045\\
BATC 6075-6660 &   -0.009  &-0.006 &  0.002 &   0.000  & 0.001 &  0.002 \\
BATC 5795-6075&     0.005  & 0.001 &  -0.004 &  0.004 & -0.001  & 0.001 \\
\Stromgren $[m_{1}]$ &      0.024  & 0.021 & -0.025 & -0.003&  -0.032  & 0.029\\
\Stromgren $[c_{1}]$  &     -0.078  &-0.015 &  0.054 &  0.003  & 0.080 & -0.009\\
\hline
\end{tabular}
\end{table}

In a non-solar scaled environment, indices may reflect those abundance
 changes. To trace the effects of different elements, we increase
the stellar atmosphere model abundances of six major elements (C, N, O, Na, Mg, Fe) by 0.3
dex\footnote{0.15 dex for C, to avoid the danger of turning to a
  carbon star \citep{ser05}. }, one single element at a time. The
sensitivity is modeled only in the stellar library; the isochrones are
kept as published.
Table \ref{tab:eh} outlines the index changes at 8 Gyr, solar metallicity.
\begin{itemize}
\item{The prominent carbon and
nitrogen effect on DDO C(41-42) colour is not surprising, since the DDO 41
band overlaps CN$_{1}$ and CN$_{2}$ Lick indices in wavelength. 
Fe and CN absorption lines are the main features under 4000 \AA,
which explains the DDO C(35-48) colour changes at super-solar C, N,
and Fe abundances. The behavior of increased O is opposite due to
molecular balancing: The CO molecule has the strongest binding energy
and so adding O decreases the C supply, weakening C$_{2}$ and CN features.
Thus, adding more O causes bluer DDO
C(41-42) and C(35-38) colours.
}
\item{The BATC 6075-6660 and 5795-6075 colours are
relatively unaffected by six major elements, which makes them good
colours to study systems with unknown element enhancements. What we find matches the
basic philosophy of creating this photometric system: avoiding known bright
feature lines and sky lines, in order to study the continuum of the
spectrum \citep{fan96}.}
\item{\Stromgren $[m_{1}]$ and $[c_{1}]$ are different from the four
    colours above, because they are the subtractions of two colours.
    $[m_{1}]$ is found to be metallicity sensitive
    in $\S$ \ref{sub:three}, which is consistent with its behavior in
    Table \ref{tab:eh}: Five elements have substantial influences on
    $[m_{1}]$.  The trend of the $[c_{1}]$ index change due to C and O abundance
    variations is reverse compared to other indices. Since $u$ and $v$
    bands are both blanketed by CN absorption lines, our models predict that
    the impact of CN on the $v$ band is greater than that on the $u$
    band. Therefore, higher CN abundance means a larger magnitude increase
    in the $v$ band than that in the $u$ band, leading to a smaller $[c_{1}]$ value. The
    strong Mg effect on $[c_{1}]$ is echoed by the Mg3835 ($u$ band) and
    Mg4780 ($b$ band) indices defined in \citet{ser05}.
}

\end{itemize}

\subsection{($B-V$) vs. ($V-K$) Plots}
\label{sub:bvk}
The possibility of breaking age-metallicity degeneracy with optical plus
near-IR colours \citep{de96, ja06, pe08}, and the upcoming era of
near-IR telescopes (eg: {\it JWST\/}) inspire us to delve
the optical near-IR colour$-$colour plot. Prior to this work, \citet{lee07} inspected the
($B-V$) vs. ($V-K$) plot using \citet[hereafter BC03]{bc03}, \citet{m05},
and BaSTI \citep{cor07} models.
They suggested different treatments of convective core overshooting and
differences in the TP-AGB phase cause the inconsistency among different
model grids (also see \citealt{m06}).
To further discuss the robustness of the models, we investigate the
isochrones that looked so promising (B94) and a later, even more complete
version (B09) from the same group.
\begin{itemize}
\item{B94 models} \\
Perhaps partly because of the inclusion of all stellar evolutionary phases,
B94 models were widely adopted as basic stellar evolution
isochrones for evolutionary synthesis models, such 
as: BC03; PEGASE \citep{fr97, fr99}; and STARBURST99
\citep{vl05} models.
Figure \ref{fig:pad94} shows the ($B-V$) vs. ($V-K$) plot of B94 models.
Three observational samples are selected for
comparison: (1) old, metal-poor GCs from \citet{bur84} 
({\it filled circles\/}); (2) old, metal-rich Es from 
\citet{pe89} ({\it filled triangles\/}); (3) SWB\footnote{\citet{SWB}} type
IV$-$VI Large Magellanic Cloud (LMC) star clusters ({\it crosses\/}). 
The reddening-corrected ($B-V$) and ($V-K$) colours come from 
\citet{van81} and \citet{per83}.
The last sample is selected as ``intermediate age'' SSPs, roughly 1
to 8 Gyr according to \citet{per83}. Generally speaking, observables
of these three samples locate at expected locations on the model grids.

\begin{figure}
\centering
\includegraphics [scale=0.4]{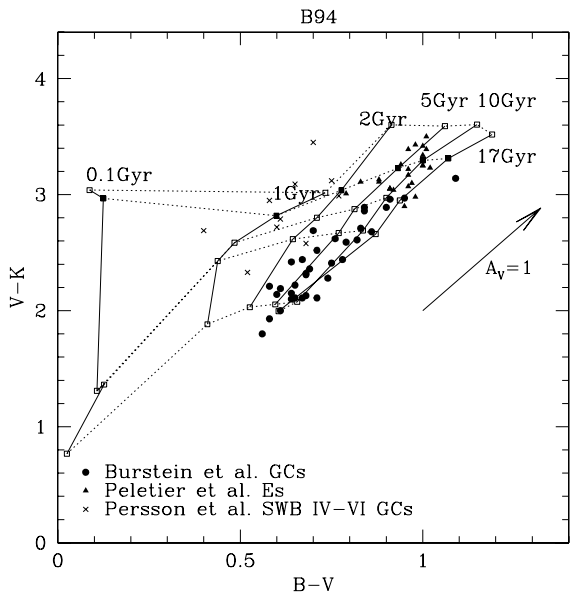} 
\caption{($B-V$) vs. ($V-K$) plot of B94 models. The models are given
  at age$=0.1, 1, 2, 5, 10, 17$ Gyr with $\log{(Z)}=-1.7, -0.7, -0.4,
0, 0.4 $. Solar metallicity SSPs are marked as {\it filled squares\/} to guide
the eye. SSPs of the same age are connected by {\it solid lines\/}, while SSPs
of the same metallicity are connected by {\it dotted
  lines\/}. A vector for A$_V=1.0$ mag is sketched to estimate the
  extinction effect. Observables from three samples are labelled 
by different symbols: {\it Filled Triangles\/}: Es from
\citet{pe89}. {\it Filled Circles\/}: GCs from \citet{bur84}. {\it Crosses\/}:
SWB type IV$-$VI Large Magellanic Cloud star clusters
from \citet{van81} and \citet{per83}.}\label{fig:pad94} 
\end{figure}

\begin{figure}
\centering
\includegraphics [scale=0.4]{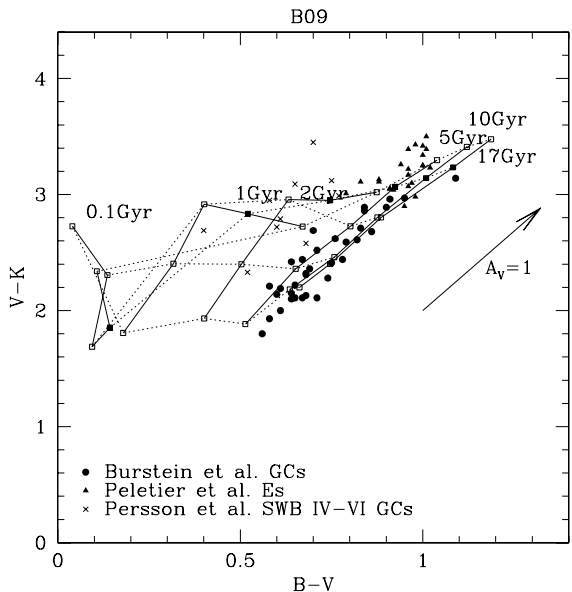} 
\caption{($B-V$) vs. ($V-K$) plot of B09 models. The models are given
  at age$=0.1, 1, 2, 5, 10, 17$ Gyr with $\log{(Z)}=-2.23, -1.23, -0.53,
0, 0.37 $. Other symbols are the same as in Figure
  \ref{fig:pad94}. }\label{fig:pad11}
\end{figure}

\item{B09 models}\\
With improved understanding of TP-AGB phase, Padova group updated
the stellar evolution models. Compared with previous
models of this group (B94; \citealt{g00}), B09 models adopted a
sophisticated TP-AGB model which includes third dredge-up, hot
bottom burning, and variable molecular opacities \citep{m08}. Also,
basic observables of AGB stars in the Magellanic Clouds (MCs) were
reproduced by the new models \citep{gm07}. 
The ($B-V$) vs. ($V-K$) plot of the B09 models are shown in Figure
\ref{fig:pad11}. Observables are also plotted as 
 in Figure \ref{fig:pad94}. Isochrones of
age larger than 5 Gyr are consistent with that of the B94 models. Es and GCs locate on the
metal-rich and metal-poor sides, respectively. However,
isochrones of 1 and 2 Gyr change substantially: the ($V-K$) colour of $\log{(Z)}=-0.53$ becomes
redder than that of solar metallicity. An unreported ``S'' shape is
found if $\log{(Z)}=-0.53$ SSPs are connected. Another unexpected change is
about the isochrone of the youngest age, 100 Myr: the ($V-K$) colour of the
lowest metallicity ($\log{(Z)}=-2.23$) is even redder that of
the highest metallicity ($\log{(Z)}=0.37$).
\end{itemize}

\section{Discussion}
\label{sect:disc}
\subsection{\Stromgren System: Empirical Leverage?}
Prior to this work, $c_{1}$ or $m_{1}$ were paired with ($u-v$), ($b-y$),
or other colours to estimate galaxy age \citep{rb89,ra96}. In this
paper, the extinction independent Q-indices $[c_{1}]$ and $[m_{1}]$
are assembled, but do a mediocre job breaking the age-metallicity
degeneracy.  
As seen in Figure \ref{fig:mc11}, several observables lie outside the B09
model grids. Three reasons are hypothesized for these outliers:
First, the conversion relations between {\it uz, vz, bz, yz\/} and
{\it uvby} system presented in \citet{ra96} are empirical, which
inevitably increases the uncertainty; Second,
the dynamic range of $[c_{1}]$ index is not large compared to the
observational error; Third, isochrones of age between 5 and 17 Gyr
are partially degenerate, which reduces the span of model grids.

\begin{figure}
\centering
\includegraphics [scale=0.4]{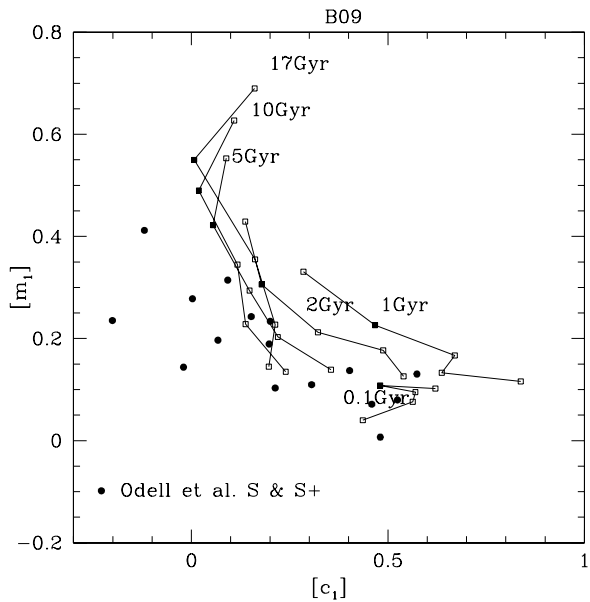} 
\caption{\Stromgren $[c_{1}]$ vs. $[m_{1}]$ plot of B09 models. The models are
  discribed in Figure \ref{fig:mc11}. Solar metallicity SSPs are marked as {\it filled squares\/} to guide
the eye. {\it Filled Circles\/}: S and S+ of \citet{o02}.}\label{fig:mcS}
\end{figure}

The model isochrone of 100 Myr lies low in the $[c_{1}]$
vs. $[m_{1}]$ plot. This is similar to the statement in \citet{ra96} that
most of the starburst galaxies have {\it mz\/}
values\footnote{$mz \equiv (vz-bz)-(bz-yz)$} less than
$-0.2$, and they are well separated
from other types of galaxies. To test the potential of $[m_{1}]$ index
as an star formation rate indicator, S
(``star formation rates equivalent to normal disk galaxy'') 
and S+(``starburst objects'') from \citet{o02} are selected and shown in Figure
\ref{fig:mcS}. Most of the  {\it mz\/} values of this sample are
less than $-0.1$, supporting the idea that starburst galaxies are
bluer in $[m_{1}]$. However, we detect no clear $[m_{1}]$ value that
separates Es and star forming galaxies.

\Stromgren system photometry is far less developed than broad band
photometry for galaxies, however. That, plus the large apparent
observational scatter in Figures \ref{fig:mceg}, \ref{fig:mc11} and
\ref{fig:mcS} lead us to conjecture that \Stromgren photometry may
still have great potential for breaking the age-metallicity
degeneracy, but higher quality data are needed to be certain.

\subsection{Model Grid Differences}
\label{sub:dif}
Three SPS models are assembled in this work: W94, B94, and B09 (i.e., 
primitive, competent, and sophisticated, respectively.). Figure
\ref{fig:mceg}, \ref{fig:mc11}, \ref{fig:pad94}, and \ref{fig:pad11}
clearly illustrate the model differences in $[c_{1}]$ vs. $[m_{1}]$ and
($B-V$) vs. ($V-K$) plots. 
During the whole work, we calculate the magnitudes of each bands in three
SPS models with the same codes, thus W94, B94, B09 models share the
same spectral library, IMF \citep{sal55} 
with lower mass limit $M_{min}=0.1 M_{\odot}$, upper mass limit $M_{max}=100
M_{\odot}$, and etc.
Most of the model grid discrepancies rely on different 
stellar evolution descriptions: 
\begin{itemize}
 \item{W94 models are a merging of isochrones and tracks from many
     sources.  Compared to B94, W94 models
have a factor of two more stars in the upper RGB. The
core-helium burning phase was approximated by a single red clump.  The
luminosities and lifetimes of AGB stars were drawn from theoretical tracks.} 
\item{B94 models included all phases of stellar evolution until the
    remnant stage in all mass ranges. Convective core overshooting
  was also considered in the models. Nevertheless, the analytic prescription
of TP-AGB phase was approximate. \citep{m08}}
\item{B09 models adopted the latest TP-AGB model of \citet{m08}.
    Compared with previous ones, this TP-AGB model has included
    several crucial effects and it was calibrated by AGB
    stars in the MCs. Bertelli et al. reckoned B09 models
    as satisfactory achievement.}
\end{itemize}

In Figure \ref{fig:pad11}, SSPs
with $\log{(Z)}=-0.53$ have the reddest ($V-K$) colours along
isochrones of 1 and 2 Gyr. This feature is not seen in B94 model
grids. Since B94 and B09 models mainly differ in TP-AGB model
treatments, which affect the near-IR colour seriously between 0.3 and 2 Gyr
\citep{m05}, it is suggested different treatments of
TP-AGB stars are responsible for the changes of model grids.
Considering the metallicities of star clusters in the LMC  (mean
$\log{(Z)}=-0.42$ for clusters with $1\le t<2$ Gyr and $-0.57$
for clusters with $2 \le t <10$ Gyr, \citealt{pe08}), SWB
IV$-$VI star clusters are better fitted in B09 models ---
B94 models give much higher metallicities.
Nevertheless, noting that B09 models are calibrated by AGB
stars from the MCs, and that the metallicities of star clusters in the
LMC are around $\log{(Z)}=-0.53$, 
it is suspected that SSPs of other metallicities,
especially the extremely low metallicity, may not be well
calibrated. The scarcity
of very low metallicity AGB stars around solar neighbourhood
impedes empirical calibration. 
It is reasonable to suggest closer
scrutiny for SSPs with extremely low metallicities around 1 and 2 Gyr.

Another unexpected feature in Figure \ref{fig:pad11} is the twist of
the 100 Myr isochrone. This youngest isochrone is unique: the ($V-K$)
colour at the lowest metallicity is redder than that of the highest
metallicity. This is inconsistent with the usual sense that metal-rich SSPs
are redder than metal-poor SSPs (See the 100 Myr isochrone of Figure
\ref{fig:pad94}). The change of the 100 Myr isochrones between B94 and
B09 models challenge the method of estimating ages and metallicities of young SSPs ($0.1<t<1$ Gyr)
from optical near-IR colour$-$colour plots \citep{bro99,hunt03}.

\begin{figure}
\centering
\includegraphics [scale=0.4]{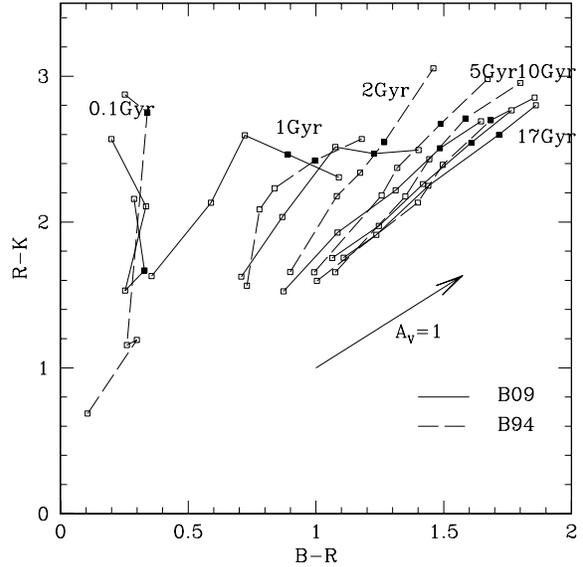} 
\caption{($B-R$) vs. ($R-K$) plot of B94 and B09 models. The B94 models are given
  at age$=0.1, 1, 2, 5, 10, 17$ Gyr with $\log{(Z)}=-1.7, -0.7, -0.4,
0, 0.4 $, while the B09 models are given
  at age$=0.1, 1, 2, 5, 10, 17$ Gyr with $\log{(Z)}=-2.23, -1.23, -0.53,
0, 0.37 $. Note that the two sets of tracks are not identical.  Isochrones
  of B09 models are labelled by {\it solid lines\/}, and isochrones of
  B94 models are labelled by {\it dashed lines}. Solar metallicity SSPs are marked as
{\it filled squares\/} to guide the eye. A vector for A$_V=1.0$ mag is sketched to estimate the
  extinction effect.
}\label{fig:brk}
\end{figure}

Finally, we check B94 and B09 model differences in the popular ($B-R$) vs. ($R-K$) plot
\citep[Figure \ref{fig:brk}]{de96,bell00}. {\it Solid lines\/} are
isochrones of B09 models; {\it dashed lines\/} are 
isochrones of B94 models. Differences between two model grids are
similar to ($B-V$) vs. ($V-K$) plots: SSPs of $\log{(Z)}=-0.53$
have the reddest ($R-K$) colours along the 1 and 2 Gyr isochrones in B09 models; 
The ($R-K$) colour at the lowest metallicity is redder than that at the
highest metallicity along the 100 Myr isochrone of B09. Thus, these features
are dependent upon near-IR band ($K$), instead of visible bands ($V$ or $R$).
Considering the TP-AGB treament differences of B94 and B09 models and the TP-AGB
effects on $K$ band, it is again suggested TP-AGB stars are responsible for the
changes of young SSP isochrones ($0.1<t<2$ Gyr), which is
consistent with \citet{m06} and \citet{lee07}

\section{Summary}
\label{sect:sum}
Motivated by the need of estimating ages and metallicities from
photometric systems, this work explored the DDO, BATC, \Stromgren
systems in the hope of finding age sensitive or metallicity sensitive
indices. Three index$-$index plots were examined but only \Stromgren $[c_{1}]$
vs. $[m_{1}]$ plot showed moderate age-metallicity separation. 
Four indices of the DDO and \Stromgren
systems changed substantially while increasing C and N abundances.
O traces an opposite trend compared to C and N, since O tends to form
CO with C.
Then, we turned to the literature-posited age-metallicity disentangling space ---
the optical plus near-IR colour$-$colour plot. 
Updated stellar evolution gives no support for clear disentanglement
of age. Even mean age, therefore, is a subtle effect in colors, and
would require, seemingly, both finer-tuned models and observations of
greater accuracy. 


\label{lastpage}

\end{document}